\documentclass[10pt,a4paper,twoside,twocolumn,american,prd,nofootinbib,raggedbottom]{revtex4}
\usepackage{lmodern}

\usepackage[T1]{fontenc}
\usepackage[utf8]{inputenc}
\usepackage{graphicx}
\usepackage{color}
\usepackage{babel}
\usepackage{booktabs}
\usepackage{units}
\usepackage{amsmath}
\usepackage{amssymb}
\usepackage{esint}
\usepackage[unicode=true,pdfusetitle,
 bookmarks=true,bookmarksnumbered=false,bookmarksopen=false,
 breaklinks=true,pdfborder={0 0 0},backref=false,colorlinks=true]
 {hyperref}
\hypersetup{
 citecolor=blue,filecolor=blue,linkcolor=blue,urlcolor=blue}

\makeatletter  

 \@ifundefined{textcolor}{}
 {%
   \definecolor{BLACK}{gray}{0}
   \definecolor{WHITE}{gray}{1}
   \definecolor{RED}{rgb}{1,0,0}
   \definecolor{GREEN}{rgb}{0,1,0}
   \definecolor{BLUE}{rgb}{0,0,1}
   \definecolor{CYAN}{cmyk}{1,0,0,0}
   \definecolor{MAGENTA}{cmyk}{0,1,0,0}
   \definecolor{YELLOW}{cmyk}{0,0,1,0}
 }

\makeatother 

\begin{document}

\title{Perturbed redshifts from N-body simulations}

\author{Julian Adamek}
\email{julian.adamek@obspm.fr}
\affiliation{Laboratoire Univers et Th\'eories, Observatoire de Paris -- PSL Research University -- CNRS -- Universit\'e Paris Diderot -- Sorbonne Paris Cit\'e,
5 place Jules Janssen, 92195 Meudon CEDEX, France}

\begin{abstract}
In order to keep pace with the increasing data quality of astronomical surveys the observed source redshift has to be modeled
beyond the well-known Doppler contribution. In this letter I want to examine the gauge issue that is often glossed over when
one assigns a perturbed redshift to simulated data generated with a Newtonian N-body code.
A careful analysis reveals the presence of a correction term that has so far been neglected. It is roughly proportional to the observed
length scale divided by the Hubble scale and therefore suppressed inside the horizon. However, on gigaparsec scales it can be comparable
to the gravitational redshift and hence amounts to an important relativistic effect.
\end{abstract}

\date{\today}

\maketitle

\section{Introduction}
\label{sec:intro}

While the standard analysis of redshift space perturbations \cite{Kaiser:1987qv} only contains the leading Doppler term, recent and future
galaxy surveys have or will have sufficient statistical power to detect subleading terms such as gravitational redshift, weak lensing,
transverse Doppler shift, time delays etc. These include general relativistic effects
that are interesting targets for testing gravity at cosmological scales. For instance, a first detection of gravitational redshift in
galaxy clusters has been claimed in \cite{Wojtak:2011ia}. Perturbation theory can be used to understand these effects analytically
(e.g.\ \cite{Bonvin:2014owa}), and many results have been derived in longitudinal gauge. However, their application to N-body simulations
requires due care with respect to how the data are mapped to the relativistic spacetime.

One possibility is to run relativistic simulations directly in longitudinal gauge \cite{Adamek:2015eda}, but the use of Newtonian N-body
codes is by far the more common practice. Coming from the Newtonian world it is not immediately clear how their results
should be interpreted in a relativistic context (e.g.\ \cite{Chisari:2011iq,Green:2011wc,Flender:2012nq}). In \cite{Fidler:2015npa} it
was finally realized that one can specify a gauge, the so-called ``N-body gauge,''\footnote{In fact, \cite{Flender:2012nq} had already
discovered ``half of'' the necessary coordinate transformation, leading to what they called ``Newtonian matter gauge.'' The spatial part
of the transformation is however different from the N-body gauge, leading to a non-vanishing volume perturbation.} in which the
relativistic equations are formally identical to the Newtonian ones, therefore providing a framework for a fully relativistic
interpretation of Newtonian simulations. The same authors went on to analyze initial conditions in this framework, showing that the usual
recipes are consistent with general relativity as well \cite{Fidler:2017ebh}.

The issue has so far mostly been discussed in relation to the matter power spectrum, and the implications for the perturbed redshift and
some other observables have therefore not yet been fully appreciated. In the recent literature the redshift formula of longitudinal gauge
is often directly applied to N-body simulation data, thereby disregarding the fact that
these are not provided in the appropriate coordinate system. For instance, \cite{Cai:2016ors,Zhu:2017} do not account for this aspect,
but I will show that the correction is fortunately very small on the scales they are interested in. While the gauge issue is noted in
\cite{Borzyszkowski:2017ayl} where a correction term is applied to the density perturbation (see their equation 17), the redshift of
individual sources is still computed without such a correction (see their equations 9 and 10). In the following I wish to clarify
this issue by studying the coordinate transformations involved, and finally by deriving the correct formulas for the perturbed redshift in
N-body gauge. I show that a small correction term due to the spatial coordinate transformation appears and should in principle be included
in the analysis. However, this term only becomes relevant on extremely large scales.

\section{Weak-field metric}
\label{sec:metric}

Astrophysical objects with high compactness exist on scales $\lesssim$~0.01~parsec (the largest known supermassive black holes),
while on extremely large scales $\gtrsim$~100~megaparsec the Universe can be described entirely in terms of linear equations. In between
these two extremes there lies a vast range of scales where the distribution of matter can be very inhomogeneous but the gravitational fields
are weak, and the geometry is therefore only weakly perturbed. This empirical fact is confirmed every time we point a telescope at the
sky and observe that most rays seem to propagate along almost straight paths.

It is therefore possible to describe the dynamics on those scales in terms of the \textit{nonlinear} evolution of matter on a
geometry with \textit{linearly} perturbed metric. Of course, as for any geometry, there also exist many coordinate
systems in which the metric takes a wildly nonlinear form. A particularly well-known coordinate system in which the smallness of the
geometric perturbation is carried into effect is the one of the longitudinal gauge. For the purpose of a more general
discussion I adopt the notation introduced in \cite{Kodama:1985bj} and write the generic line element for a metric with linear scalar
perturbations on top of a Friedmann-Lema\^itre model as
\begin{multline}
 ds^2 = a^2(\tau) \biggl[-\left(1 + 2 A\right) d\tau^2 + \left(1 + 2 H_L\right) \delta_{ij} dx^i dx^j\\- 2 \nabla_i B dx^i d\tau - 2 \left(\!\nabla_i \nabla_j - \frac{1}{3} \delta_{ij} \Delta\!\right)\! H_T dx^i dx^j\biggr]\,,
\end{multline}
where $a(\tau)$ is the scale factor, $\tau$ and $x^i$ are conformal time and comoving coordinates on the spacelike hypersurface,
respectively, and $A$, $B$, $H_L$, $H_T$ are scalar functions describing perturbations.
Vector and tensor perturbations shall not be discussed here. A setting is said to be ``weak-field'' if there exists a coordinate system
for which all the above perturbation variables are small, i.e.\ $|A|$, $|H_L|$, $\sqrt{\nabla_i B \nabla^i B}$, $\sqrt{\nabla_i \nabla_j H_T \nabla^i \nabla^j H_T - (\Delta H_T)^2/3} \ll 1$. One can then find other such coordinate systems by making a small change of coordinates,
generated by two scalar fields $T$ and $L$, so that $\tau \rightarrow \tau + T$ and $x^i \rightarrow x^i + \nabla^i L$. This shows that
the four scalar perturbations only contain two physical modes. The longitudinal gauge is obtained by choosing $T$
and $L$ such that $B = H_T = 0$, leaving only the lapse perturbation $A = \Psi$ and the volume perturbation $H_L = \Phi$, where the two
potentials $\Psi$ and $\Phi$ denote precisely the two physical modes when written as first-order gauge-invariant expressions. The typical
amplitude of these perturbations is $\sim 10^{-5}$ in our Universe, except for the vicinity of black holes or neutron stars where the
weak-field description breaks down.

In longitudinal gauge the Hamiltonian constraint reads
\begin{equation}
 -\Delta \Phi + 3 \mathcal{H} \Phi' - 3 \mathcal{H}^2 \Psi = 4 \pi G a^2 \delta\rho\,,
\end{equation}
where $\mathcal{H}$ is the conformal Hubble rate, a prime denotes partial derivative with respect to $\tau$, and the equation is
linearized in $\Phi$ and $\Psi$ but not in $\delta\rho$. In fact, the matter perturbation $\delta\rho$ can be very large, but it is
computed and evolved on a linearly perturbed geometry. This equation is however not the one used in a Newtonian N-body code. First, such
a code uses counting densities that are not corrected for perturbations of the volume element -- Newtonian theory assumes Euclidean
geometry. Second, the Poisson equation lacks some of the terms featured on the left-hand side.

From now on I restrict the discussion to the case where gravitational fields are sourced exclusively by nonrelativistic matter.
In this case anisotropic stress can be neglected, implying $\Phi = -\Psi$.

\section{Perturbed redshift in longitudinal gauge}
\label{sec:longitudinal}

The effect of geometry (and perturbations thereof) on observables can be understood by studying the geodesics of photons that reach
the observation event. Let me denote the tangent vector of such a geodesic as $k^\mu$.
The condition $g_{\mu\nu} k^\mu k^\nu = 0$ implies that $k^i/k^0 = n^i (1 + 2 \Psi)$, where $n^i$ is the unit
vector ($\delta_{ij} n^i n^j = 1$) pointing in the direction the photon is traveling. The geodesic equation can then be expressed
as an evolution equation for the energy,
\begin{equation}
\label{eq:k0long}
 \frac{d\ln k^0}{d\tau} = - 2 \mathcal{H} - 2 n^i \nabla_i \Psi\,,
\end{equation}
and an equation describing the deflection of the ray,
\begin{equation}
 \frac{dn^i}{d\tau} = -2 \left(\delta^{ij} - n^i n^j\right) \nabla_j \Psi\,.
\end{equation}
A reference clock is specified through a unit timelike vector
$u^\mu = a^{-1} (1-\Psi, v^i)$ where $v^i = dx^i / d\tau$ is the peculiar (coordinate) velocity of the clock's rest frame.
In such a frame, the measured photon energy is $-g_{\mu\nu} u^\mu k^\nu$. A first integral of
eq.~(\ref{eq:k0long}) yields the following expression for the observed redshift between source (src) and observer (obs):
\vspace{-0.5\baselineskip}
\begin{multline}
\label{eq:zlong}
 1+z = \frac{g_{\mu\nu} u^\mu k^\nu|_\mathrm{src}}{g_{\mu\nu} u^\mu k^\nu|_\mathrm{obs}} = \\\frac{a_\mathrm{obs}}{a_\mathrm{src}} \Biggl(1 + n_i v^i_\mathrm{obs} - n_i v^i_\mathrm{src} + \Psi_\mathrm{obs} - \Psi_\mathrm{src} - 2\!\int\limits_\mathrm{src}^\mathrm{obs}\!\Psi' d\chi\Biggr)
\end{multline}
Here $d\chi$ is a conformal distance element along the photon path. The redshift perturbations are easily identified as the Doppler
shift due to peculiar motion, the gravitational redshift due to time dilation, and the Rees-Sciama effect (also known as integrated
Sachs-Wolfe effect in the context of linear theory).

The boundary terms in this expression have to be evaluated at the coordinate time at which the photon geodesic actually intersects the
world lines of source and observer. The time of flight of the photon is affected by the Shapiro delay, which for a coordinate distance
$\chi$ between source and observer is given by the following relation:
\begin{equation}
\label{eq:Shapiro}
 \chi = \!\int\limits_\mathrm{src}^\mathrm{obs}\! d\chi = \!\int\limits_\mathrm{src}^\mathrm{obs}\! \left(1 + 2 \Psi\right) d\tau = \tau_\mathrm{obs} - \tau_\mathrm{src} + 2 \!\int\limits_\mathrm{src}^\mathrm{obs}\! \Psi d\chi
\end{equation}

\section{From longitudinal to N-body gauge}
\label{sec:N-body}

Given a linearly perturbed geometry specified in longitudinal gauge I now make a small change of coordinates to set
$H_L = 0$. Considering the Lie derivative of the metric tensor the required transformation is generated by $T$, $L$
that satisfy
\begin{equation}
\label{eq:volume}
 \mathcal{H} T + \frac{1}{3} \Delta L = \Psi\,.
\end{equation}
I furthermore choose $L' = 0$ such that velocities are not transformed.
In this new coordinate system one finds $B = T$ and $H_T = -L$. It is important to verify that these new perturbations do not
violate the weak-field conditions\footnote{For instance, had I made a change of coordinates that sets $A = B = 0$ instead, like in
a synchronous gauge, I would have found $L' = T = -a^{-1}\int^a (\Psi/\mathcal{H})d\tilde{a}$ and that $H_L$ receives a contribution
$\sim \Delta\Psi/\mathcal{H}^2 \sim \delta\rho/\rho$. The volume perturbation therefore does not remain small everywhere in such a
coordinate system, rendering the weak-field treatment inconsistent.}. As explained below, $L$ shall be chosen such that $\Delta L$
is of order $\Psi$, and hence the same is true for the perturbations generated by $H_T$. One can then easily convince oneself that
$\nabla_i T$ is of the order of a peculiar velocity --- in fact, it coincides with the Zel'dovich approximation thereof --- and therefore
$\sqrt{\nabla_i B \nabla^i B} \sim v \lesssim 10^{-3}$ at low redshift.
The shift perturbation in the new coordinate system is therefore substantially larger than $\Psi$, but still comfortably
within the weak-field regime.
The lapse perturbation becomes
\begin{equation}
\label{eq:lapse}
 A = \Psi + \mathcal{H} T + T'\,,
\end{equation}
and as explained shortly I arrange that it vanishes at leading order.

The Hamiltonian constraint becomes
\begin{equation}
\label{eq:Poisson}
 \Delta\!\left(\!\mathcal{H}B - \frac{1}{3} \Delta H_T\!\right) + 3 \mathcal{H}^2 A = 4 \pi G a^2 \delta\rho\,,
\end{equation}
and as a consequence of $H_L = 0$ the density perturbation $\delta \rho$ can be obtained simply by counting the mass elements per
coordinate volume, in accordance with the procedure relevant for Newtonian codes.
For nonrelativistic particles the geodesic equation
reduces to
\begin{equation}
\label{eq:acc}
 \frac{dv^i}{d\tau} + \mathcal{H} v^i = \nabla^i\left(\mathcal{H} B + B' - A\right)\,.
\end{equation}
Since I assume that matter is nonrelativistic and hence
the pressure perturbation can be neglected, the spatial trace of Einstein's equations yields
\begin{equation}
\label{eq:trace}
 \mathcal{H} A' - \left(\mathcal{H}^2 - 2\frac{a''}{a}\right) A = 0 \quad \Rightarrow \quad A \propto \frac{1}{a\mathcal{H}^2}\,.
\end{equation}
This shows that an appropriate choice of boundary conditions will set $A = 0$.
Furthermore, with such a choice eqs.~(\ref{eq:Poisson}), (\ref{eq:acc}) are formally identical to the ones of Newtonian
gravity if eqs.~(\ref{eq:volume}), (\ref{eq:lapse}) are used,
\begin{equation}
 \Delta \Psi = 4 \pi G a^2 \delta\rho\,,\qquad \frac{dv^i}{d\tau} + \mathcal{H} v^i = -\nabla^i \Psi\,.
\end{equation}

Keeping in mind that the above equations remain valid even if $\delta\rho/\rho$ becomes large I now want to set the boundary conditions
at early times when matter perturbations are still linear. According to eq.~(\ref{eq:lapse}) the condition $A = 0$ is satisfied when
\begin{equation}
 T = -\frac{1}{a}\int\limits^a \!\frac{\Psi}{\mathcal{H}} d\tilde{a}\,.
\end{equation}
The linear solution of $\Psi$ is constant in matter domination, and with
eq.~(\ref{eq:volume}) the corresponding choice of $L$ is given by $\Delta L = 5 \Psi_\mathrm{in}$. Here I introduce $\Psi_\mathrm{in}$
to denote the linear initial condition for $\Psi$ in matter domination.

With this choice one can see that $\nabla^i B = v^i$ in the linear regime, which (together with $H_L = 0$) is the original gauge condition
used in \cite{Fidler:2015npa} for the N-body gauge. So even though my gauge condition $H_T' = -L' = 0$ is different, the resulting
coordinate system is the same whenever only nonrelativistic matter is present. The advantage of my condition is that it does not
explicitly refer to a matter perturbation and can hence be easily extended into the nonlinear regime of matter as long as gravitational
fields remain weak. In a slight abuse of terminology I shall therefore always call this system of coordinates the one of N-body gauge, as
the state of a Newtonian N-body simulation is given in precisely these coordinates.

I shall now discuss the repercussions of this change of coordinates from longitudinal to N-body gauge. The null condition is solved in
N-body gauge by $k^i/k^0 = n^i + \nabla^i B + n^j (\nabla_j \nabla^i - \delta^i_j \Delta/3)H_T$, and the photon geodesic
equation can be written as
\begin{equation}
\label{eq:k0Nb}
 \frac{d\ln k^0}{d\tau} = -2\mathcal{H} - n^i n^j \nabla_i \nabla_j B\,,
\end{equation}
and
\begin{equation}
 \frac{dn^i}{d\tau} = -\!\left(\delta^{ij} - n^i n^j\right)\! \nabla_j \!\left(\!n^k \nabla_k B - \frac{1}{3} \Delta H_T\!\right)\!\,.
\end{equation}
Considering how the coordinate transformation acts on $u^\mu$ one sees that $u^\mu = a^{-1} (1, v^i)$ in the new coordinates. Thus,
a first integral of eq.~(\ref{eq:k0Nb}) gives the following new expression for the observed redshift:
\begin{multline}
\label{eq:Nbredshift}
 z+1 = \frac{a_\mathrm{obs}}{a_\mathrm{src}} \Biggl(1 + n_i v^i_\mathrm{obs} - n_i v^i_\mathrm{src} + \Psi_\mathrm{obs} - \Psi_\mathrm{src}\\
 +\mathcal{H}B\vert_\mathrm{obs} - \mathcal{H}B\vert_\mathrm{src} - 2\!\int\limits_\mathrm{src}^\mathrm{obs}\!\Psi' d\chi\Biggr)
\end{multline}
In order to recover all the terms of eq.~(\ref{eq:zlong}) I used $(\mathcal{H} B)' = \Psi'$ and $\mathcal{H}B + B' = -\Psi$, but
evidently a new boundary term $\mathcal{H}B$ appears. Noting that $B = T$ this boundary term can be understood as the
result of the change of coordinates acting on the background term $a_\mathrm{obs}/a_\mathrm{src}$. In other words, the term has to appear
because the equal-time hypersurfaces in N-body gauge do not coincide with the ones of longitudinal gauge. The coordinate time
of a Newtonian N-body simulation is the one of N-body gauge, and hence this boundary term needs to be taken into account.

Let me now inspect the time of flight for the photon in N-body gauge,
\begin{multline}
\label{eq:NbShapiro}
 \chi = \!\int\limits_\mathrm{src}^\mathrm{obs}\! \left(1 + n^i \nabla_i B + n^i n^j \nabla_i \nabla_j H_T - \frac{1}{3} \Delta H_T\right)d\tau\\
 = \tau_\mathrm{obs} - \tau_\mathrm{src} + B_\mathrm{obs} - B_\mathrm{src} + n^i \nabla_i H_T \vert_\mathrm{obs} \\- n^i \nabla_i H_T \vert_\mathrm{src} + 2 \!\int\limits_\mathrm{src}^\mathrm{obs}\! \Psi d\chi\,,
\end{multline}
where I again use the gauge conditions to recover the terms known from longitudinal gauge. Compared to eq.~(\ref{eq:Shapiro}) there are
two new boundary terms. These are expected from the gauge transformation, since the coordinate time transforms as $\tau \rightarrow \tau + T = \tau + B$, and the coordinate distance transforms as $\chi \rightarrow \chi + n^i \nabla_i L\vert_\mathrm{obs} - n^i \nabla_i L\vert_\mathrm{src} = \chi - n^i \nabla_i H_T\vert_\mathrm{obs} + n^i \nabla_i H_T\vert_\mathrm{src}$.
For a photon trajectory with fixed endpoints, the coordinate time of emission and observation therefore transforms such that the
boundary terms due to the shift perturbation in eqs.~(\ref{eq:Nbredshift}) and (\ref{eq:NbShapiro}) cancel exactly.
The other boundary term in eq.~(\ref{eq:NbShapiro}) gives precisely the change in the coordinate distance due to the spatial transformation
between longitudinal and N-body gauge. Therefore, the perturbed redshift for the trajectory remains invariant.

\section{Discussion}
\label{sec:discussion}

The precedent analysis clarifies that in order to use the longitudinal gauge for computing the perturbed redshifts with N-body simulation
data one should transform the coordinates appropriately. In particular, the coordinate distance between observer and sources changes
according to a spatial transformation that is independent of time. This has already been pointed out in \cite{Chisari:2011iq}, and I
explicitly show how to recover this result in the relativistic framework provided by the N-body gauge. Alternatively the computation of
the perturbed redshift can also be carried out directly in N-body gauge. In this case the coordinates of sources are directly taken from
the simulation, and the effect appears as a modification of the Shapiro delay.

In order to estimate the amplitude of the correction, let me compute the typical change $\delta\chi = n^i \nabla_i L\vert_\mathrm{obs} - n^i \nabla_i L\vert_\mathrm{src}$ of the coordinate distance. Using the relation $\Delta L = 5 \Psi_\mathrm{in}$ the variance of $\delta\chi$
is given by
\begin{equation}
 \left\langle \delta \chi^2\right\rangle = 50 \!\int\limits_0^\infty\!\frac{dk}{k^3}\!\left[\frac{1}{3} - \frac{1}{k \chi}j_1(k\chi) + j_2(k\chi)\right]\!\Delta_k^\Psi \,,
\end{equation}
where $\Delta_k^\Psi$ is the dimensionless power spectrum of $\Psi_\mathrm{in}$. Unfortunately the integral has an infrared divergence
for nearly scale invariant spectra, but in practice this divergence is regulated by the finite size of a simulation. Imposing a cutoff
close to the Hubble scale one finds that $\delta\chi / \chi \sim 10^{-4}$ almost independent of scale $\chi$ and precise value of the
cutoff. Considering how a change in the coordinate distance affects the time of flight one sees that the typical correction to
the redshift due to this coordinate effect is $\delta z / (1 + z) \sim \mathcal{H}\delta \chi$. Therefore the effect becomes of the order
of the gravitational redshift for trajectories at or above the gigaparsec. At these extreme scales the gauge correction is of the same
order as all other relevant terms and should be taken into account.

As suggested in \cite{Wojtak:2011ia} the gravitational redshift may be measured statistically by looking for ``excess redshift'' of
the brightest galaxies at the center of clusters when compared to the fainter galaxies in the outskirts. For simulating such a measurement
the relevant correction is given by the difference in time of flight between the different sources, and therefore the effect is suppressed
by the small ratio between the scale of the cluster and the Hubble scale. One expects that in this case the correction is typically less
than $1\%$ of the signal and can therefore safely be neglected.


\begin{acknowledgments}
\vspace{-0.5\baselineskip}
I thank R Durrer, C Rampf and Y Rasera for comments on the manuscript as well as C Fidler and T~Tram for many insightful discussions about 
the N-body gauge. I further enjoyed a valuable correspondence with D~Bertacca and C Porciani about \cite{Borzyszkowski:2017ayl}. Prior to
submission, C Fidler et al.\ also kindly shared their manuscript \cite{Fidler:2017pnb} on a closely related topic.
\end{acknowledgments}

\bibliographystyle{utcaps}
\bibliography{nbgauge}

\end{document}